\documentclass[twocolumn,showpacs,prl,aps,10pt]{revtex4-1}

\usepackage{graphicx} 
\usepackage{color}
\usepackage{amsmath}
\usepackage{amssymb}
\usepackage{wasysym}

\begin{document}

\title{Fractographic aspects of crack branching instability using a phase field model}

\author{H. Henry$^{1}$ and M. Adda-Bedia$^{2}$}
\affiliation{$^{1}$Physique de la Mati\`ere Condens\'ee, Ecole Polytechnique, CNRS, 91128 Palaiseau, France\\
$^{2}$Laboratoire de Physique Statistique, Ecole Normale Sup\'erieure, UPMC Paris 6, Universit\'e Paris Diderot, CNRS, 24 rue Lhomond, 75005 Paris, France}

\begin{abstract}
A phase field model of a crack front propagating in a three dimensional brittle material is used to study the fractographic patterns induced by the branching instability. The numerical results of this model give rise to crack surfaces that are similar to those obtained in various experimental situations. Depending on applied loading configurations and initial conditions, we show that the branching instability is either restricted to a portion of the crack front or revealed through quasi two dimensional branches. For the former, the crack front leaves on the main broken surface either aligned or disordered parabolic marks. For the latter, fractography reveals the so called {\em \'echelons} cracks showing that branching instability can also induce crack front fragmentation.
\end{abstract}

\date{\today}


\maketitle

Since the pioneering work of Griffith~\cite{Griffith}, there has been a dramatic progress in the understanding of how and where a crack nucleates and propagates in an initially unbroken solid~\cite{Freund90,Fineberg99}. Nonetheless, some fundamental questions remain open such as the physical mechanisms leading to 
mist and hackle crack surfaces~\cite{Bouchaud02,Ravi98,Hull99} or to branching
instability~\cite{Ravi84a,Fineberg91,Sharon1995,Sharon1996,Sharon1996a,Sagi04}.
For the former there are some hints supported by experiments~: nucleation of
micro cracks ahead of the crack
front~\cite{Ravi84b,Ravi97,Scheibert10,Guerra12}  or \textit{micro-branching}
instabilities~\cite{Ravi84a,Sharon1996a}. For the latter, studies in the
framework of Linear Elasticity Fracture Mechanics (LEFM) based on both symmetry
considerations and energetic criteria have shown that the branching instability
cannot occur below a threshold speed of the crack front~\cite{Adda05,Katzav07}.
However, these studies do not allow to establish a  mechanism for the branching
instability itself. This is presumably because in LEFM, the process zone in
which the dissipation induced by the crack propagation takes place is reduced
to a line. As a result any instability mechanism occurring there cannot be
predicted~\cite{Bouchbinder10}. Numerical tools that introduce a spatial
extension of the process zone have  been developped  to overcome this issue.
Among them, the phase field model~\cite{Karma01,Aranson00} has shown its
ability to properly describe crack propagation in various quasi-static and
dynamic
situations~\cite{Henry04,Henry08,Karma04,Spatschek06,Spatschek10,Hakim09}.
{However, it has not been yet confronted  directly to experimental results of
dynamic fracture which are primarily 3D.}
 
The  phase field model of crack propagation is a phenomenological model that describes the growth of a crack as a phase transition between a purely elastic solid and an infinitely soft phase~\cite{Karma01,Aranson00}. The evolution equation of the phase field is governed by a Ginzburg-Landau type free energy coupled to the stress field through the elastic energy density. This approach can be related to $\Gamma$-convergence theory and \textit{non local} damage theory~\cite{Bourdin00}. It has been proven to retrieve many aspects of LEFM for both two~\cite{Henry04,Henry08,Karma04,Spatschek06,Spatschek10,Hakim09} and three dimensionnal crack propagation~\cite{Pons10,Henry10}. For instance, it reproduces the theoretical prediction of the onset of the branching~\cite{Karma04,Henry08} and the instability of a single crack front under mixed mode (I+III) loading~\cite{Pons10}. In this work, this model is used to simulate three dimensional instabilities of fast cracks in a brittle material~\cite{Henry10} and  compare the results to the experimental fractographic patterns~\cite{Ravi98,Ravi84a,Fineberg91,Sharon1995,Sharon1996,Sharon1996a,Sagi04,Ravi84b,Ravi97,Scheibert10,Guerra12}. We first recall briefly the phase field model and describe the numerical setup. Thereafter, the qualitative behaviour of fracture is presented as the propagation speed is increased. The results show that the three dimensional branching instability can lead to many fractographic patterns observed experimentally, such as echelon cracks~\cite{Sagi04} or periodic crescent marks~\cite{Ravi97,Scheibert10}.

The phase field model relies on the introduction of an auxiliary field
$\varphi$ that varies between 0 and 1 and is coupled to the elastic field such
that the material is infinitely soft when $\varphi=0$ and obeys the laws of
linear elasticity in the regions where $\varphi=1$. The equations for $\varphi$
and the elastic field derive from the following  free energy density~:
\begin{equation}
	\label{eq_free_energy}
		 \mathcal{F}=\frac{D}{2}|\mathbf{\nabla} \varphi|^2+h
		 V(\varphi)+g(\varphi)\left[\frac{\lambda}{2} \varepsilon_{ii}^2+\mu
		 \varepsilon_{ij}^2 -\varepsilon_c\right]\,,
\end{equation}
where $\varepsilon_{ij}$ ($i=x,\! y,\!z$) is the strain tensor, $\lambda$ and $\mu$ are the Lam\'e coefficients, $D$, $h$ and $\varepsilon_c$ are model parameters that govern the phase field interface width, its energy and the phenomenological behaviour of the model. The function $V(\varphi)=\varphi^2(1-\varphi)^2$ is a double well potential and $g(\varphi)=4\varphi^3-3\varphi^4$ is a coupling function chosen so that the equilibrium configuration of the one dimensional crack problem  is reached when the stress is completely relaxed in the unbroken material~\cite{Karma04}. The evolution equations of both the displacement field $u_i$ and $\varphi$ write
 \begin{eqnarray}
	 \rho \partial_{tt} u_i &=&  \frac{\partial}{\partial x_j}\frac{\partial \mathcal{F}}{\partial \varepsilon_{ij}}\,,\label{eq_onde}
	 \\
	 \tau \partial_{t} \varphi &=& \frac{\partial}{\partial x_i}\frac{\partial \mathcal{F}}{\partial( \nabla_i\varphi)}-\frac{\partial \mathcal{F}}{\partial \varphi}\;.\label{eq_ev_phi}
\end{eqnarray}
The evolution equation of the displacement field conserves the total (kinetic
plus free) energy while the evolution equation of the phase field guarantees
that $\mathcal{F}$ decreases with time. Notice that the kinetic coefficient $\tau$ is a
measure of the energy dissipation at the crack front
and the  the fracture energy  at zero velocity is given by $G_0=\sqrt{2D}\int_0^1
d\varphi \sqrt{hV(\varphi)+(1-g(\varphi))\varepsilon_c}$~ \cite{Karma04,Henry08}.
Finally, the resulting width of the phase field interface $w_\varphi$ can be seen as the spatial extension of a process zone.

\begin{figure}
\includegraphics[width=0.4\textwidth]{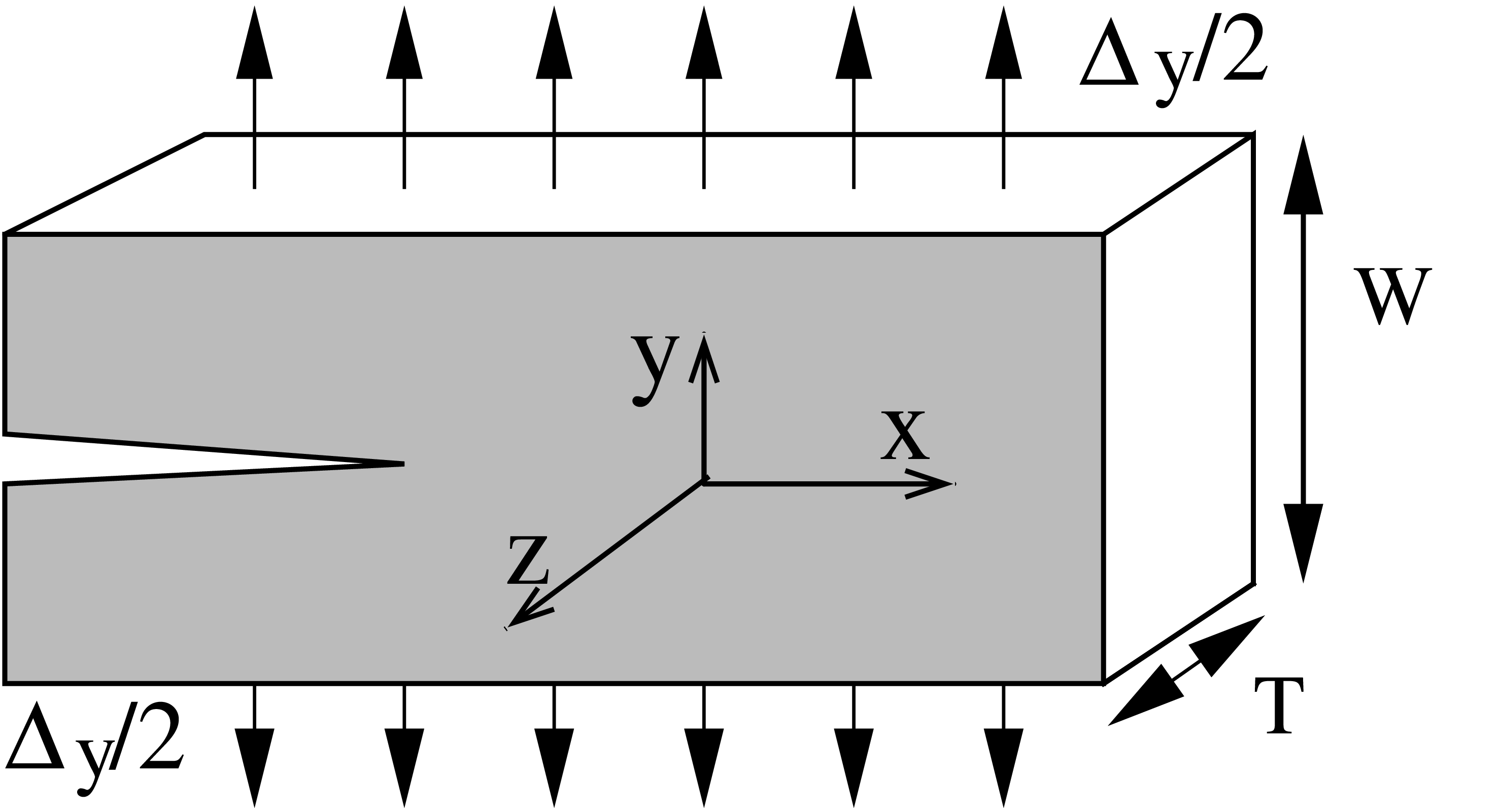}
\caption{The simulated set up. A crack front extending along the $z$-axis is propagating along the $x$-axis. Loads are applied through uniform displacements $\pm\Delta_y/2$ at $y=\pm W/2$.}
\label{fig_setup}
\end{figure}

In the following, the evolution equations (\ref{eq_onde},\ref{eq_ev_phi}) are
applied to an infinitely long parallelepiped of linear elastic material subject
to a mode~I loading (see Fig.~\ref{fig_setup}). The simulations are performed
on a sample of size $W=160$ and $T=120$, density $\rho=1$ and Lam\'e
coefficients $\lambda=\mu=1$ corresponding to a Poisson ratio $\nu=0.25$ and a
shear wave speed $c_s=1$. The following phase field parameters are used~:
$D=2$ , $h=1$, $\varepsilon_c=1$ and $\tau=2$ which corresponds to a low
dissipation at the crack front~\cite{Henry08}. Using these values, one finds
$w_\varphi\approx2$ which  is two orders of magnitude smaller  than the sizes of the sample $W$
and $T$.  Additional simulations were performed using $D=4$, $h=1/2$ and
$\varepsilon_c=1/2$ which keeps  fracture energy constant, while $w_\varphi$ is 4 times larger. The results
display no qualitative changes indicating that  scale separation is large
enough.
 
The equations were simulated on a regular tridimensional grid of size $L\times
W\times T$ moving along the $x$-axis. As an initial condition, the material was
pre-broken through a half-plane whose front was slightly perturbed by a sine
wave with an amplitude of a few grid points. The crack dynamics is controlled
by the imposed displacements $\pm\Delta_y/2$ at $y=\pm W/2$ that determine also
the elastic energy  per unit surface in the $xz$-plane stored in the
intact material, $G=(\lambda/2+\mu)\Delta_y^2/W$. At the planes $z=\pm T/2$, either periodic or surface-free
boundary conditions were used. While the latter breaks translational invariance
along the $z$-axis and prevents purely bi-dimensional pattern to appear, the
former induces a self-interaction of the crack front with itself. 
To follow cracks over long distances the simulation window  was regularly
shifted by one grid point in the $x$-direction keeping the  most advanced point
of the  crack front close  to the center of the grid~\cite{Karma04}. At
$x=-L/2$, displacements were inherited from the previous grid shift and at
$x=L/2$, they were  $u_x=u_z=0$ and $u_y=y\Delta/W$. To avoid wave reflection
at $x=L/2$, damping  in Eq.~\ref{eq_onde} was introduced close to this boundary through a
$-\eta\dot{\mathbf{u}}$ term. We checked that this does
not affect the crack dynamics by considering systems with different lengths $L$
($2W$ and $4W$). During  simulations, the position of the crack front and  the
shape of the crack surface were recorded.

Let us now turn to the description of  numerical results. At low crack speeds,
the crack front propagates at constant speed by keeping its location in the plane $y=0$. At least for the  parameters used here, the crack front does not exhibit any dynamic instability and the crack surfaces are always mirror-like. When the crack speed is increased above a threshold value that depends on
dissipation and model parameters~\cite{Henry08}, the propagation of a single
crack becomes unstable both in the two-dimensional and three-dimensional
cases~: through a tip splitting instability, the crack front branches into two
cracks that further propagate until the speed of one crack becomes faster and
screens the slower one~\cite{Karma04,Henry08}. In the three-dimensional case,
the nature of the instability is similar but the translational invariance along
the crack front is broken, leading to various patterns that are described below. One should note that while in two dimensions, the threshold crack tip speed for branching is well determined, the local character of the instability prevents its accurate computation in the three dimensional case. Before analyzing the fractograpgy of the broken surface induced by branching instability, the single branching event will be described in detail.

\begin{figure}
\includegraphics[width=0.4\textwidth]{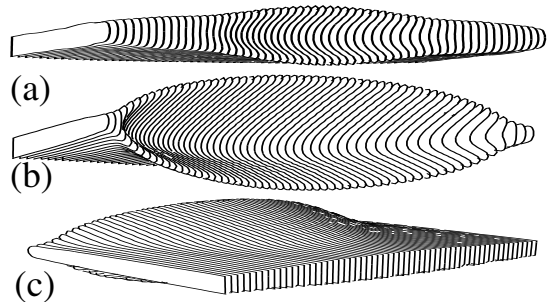}
\caption{Perspective view of the iso-surface $\varphi=0.5$ (that can be considered as the crack surface) at different times of a branching event for a thin system ($T\approx 12 w_\varphi$) with traction-free boundary conditions at $z=\pm T/2$. \textbf{(a)} A tip splitting instability is initiated at the middle of the crack front and \textbf{(b)} spreads along the crack front. \textbf{(c)} represents the same state as \textbf{(b)} seen from the rear crack front.}
\label{figlocalbranching}
\end{figure}

Fig.~\ref{figlocalbranching} shows the development of a branching event in a
thin plate ($T\approx 12 w_\varphi$) slightly above threshold. In this case,
the crack front starts to branch locally in the $z$-direction, then the region
where splitting occurs grows along the crack front until one of the branches
stops and a single crack propagation regime resumes. While this behaviour is
not surprising in the case of a sample of finite thickness, it is also observed
in the case of periodic boundary conditions long after the initial
branching event.    

\begin{figure}
\includegraphics[width=0.48\textwidth]{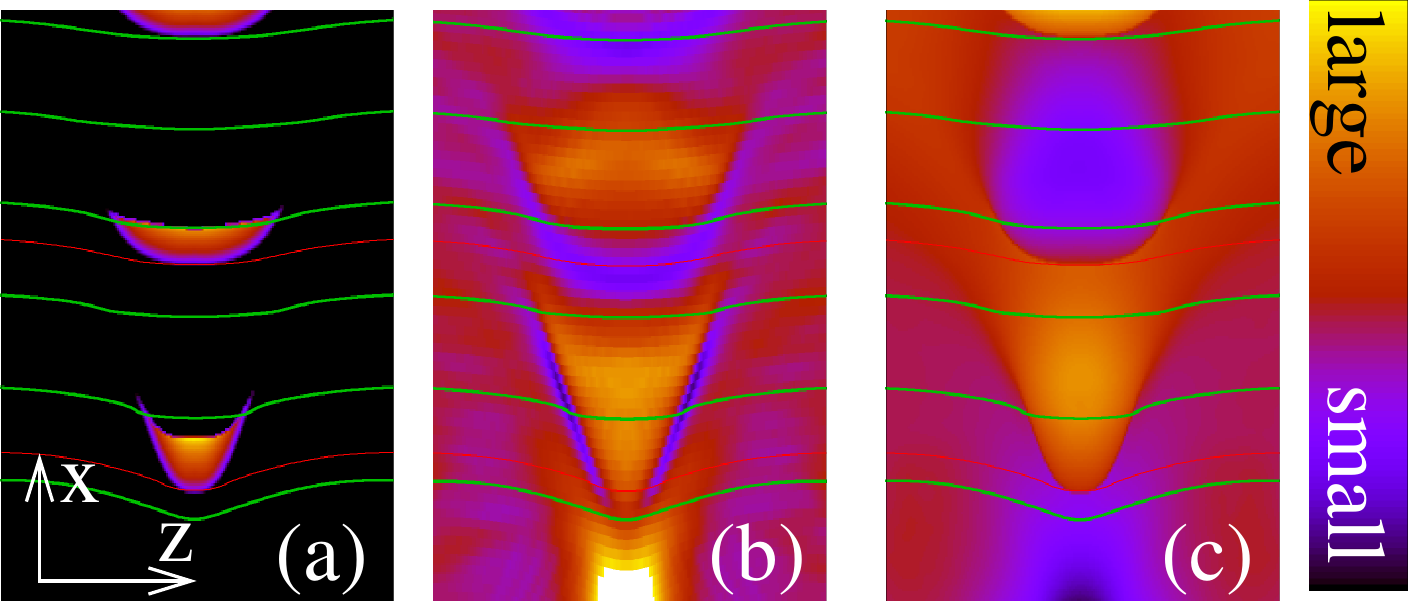}
\caption{$xz$ contour plots of crack front dynamics and fractography of the resulting surface near the instability threshold. A window of the total fracture surface ($0.2\,T$ in the $z$-direction and $0.375\,T$ in the $x$-direction) is shown. In \textbf{(a)} branching events are displayed~: black region corresponds to a single crack front phase. Elsewhere the distance in the $y$-direction between the two propagating fronts is shown. In \textbf{(b)} the local instantaneous crack front velocity of the most advanced front is plotted showing that the crack front dynamics is correlated with branching instability. The crack front slows down locally before nucleation of a branching event (purple V marks) and accelerates after the arrest of the secondary branch. \textbf{(c)} Shape of the post-mortem fracture surface. The height of the crack surface oscillates around the mean position $y=0$. In all figures, lines display instantaneous positions of the front in the single crack propagation phase.} 
\label{fig_alley}
\end{figure}

For samples of larger thickness, the branching instability also occurs locally
on the crack front and the successive branching events follow the  scenario
described above. Fig.~\ref{fig_alley} shows a typical example of this situation
in which branches do not spread through the whole $z$-direction and tend to
align along the main direction of crack propagation. This branching pattern is
reminiscent of experimental observations where a roughly periodic structure,
which is approximately in phase with the crack velocity oscillations, is
formed~\cite{Sharon1995,Sharon1996,Sharon1996a}. Our  results allow to
correlate branching instability with local instantaneous velocity of the crack
front and give some rationale for the existence of aligned branching events.
Fig.~\ref{fig_alley} shows that once a branching event has occurred, the crack
front slows down at the branching point while the rest of the front is not
perturbed. The subsequent evolution leads to the formation of a cusped V-shape
which accelerates after the branching event is finished in order to recover a
flat crack front. This implies that the local speed is more likely above the
threshold velocity for branching and thus the next branching event occurs at
the same position along the crack front. The propagation of elastic waves
introduces some randomness in the system and  prevents this deterministic branching scheme to repeat indefinitely. Fig.~\ref{fig_alley}c shows that each branching event gives rise to a vertical deformation of the main crack surface that persists over a distance longer than the branching event itself. The induced pattern on the crack surface is similar to parabolic marks observed in experiments~\cite{Ravi84b,Ravi97,Scheibert10,Guerra12} with the significant difference that the phase field simulations do not resolve scales smaller than $w_\varphi$ while experimental observations are made at the actual scale of the process zone. Nevertheless, the question whether this similarity persists when $w_\varphi$ is diminished by a few orders of magnitude remains open. 

\begin{figure}
\includegraphics[width=0.48\textwidth]{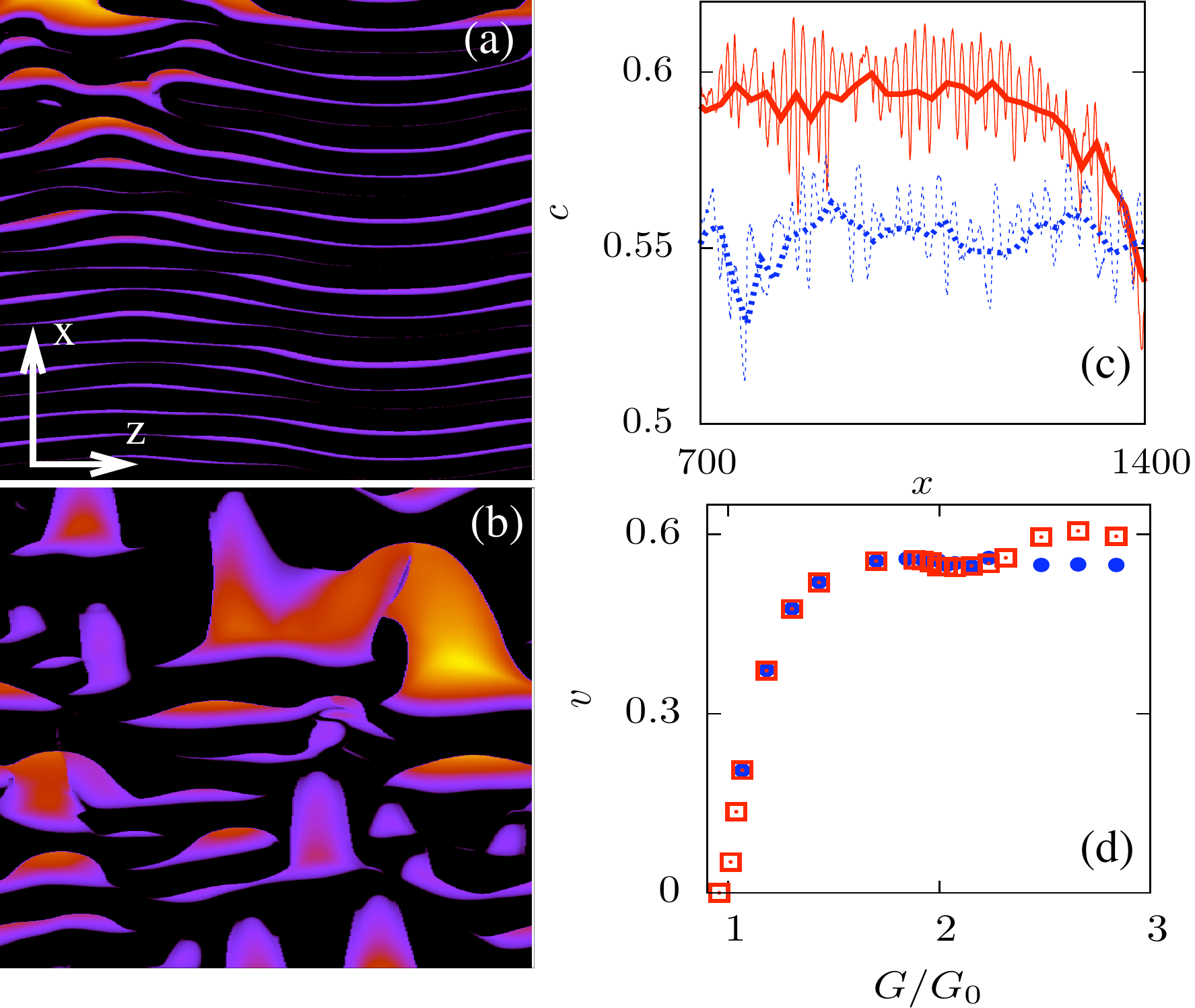}
\caption{\textbf{(a,b)}--Two fracture morphologies obtained using different initial perturbations of the crack front. The colour plots use the same representation of branching as in Fig.~\ref{fig_alley}a. In \textbf{(a)} branching spreads through the thickness of the sample while in \textbf{(b)} there are localized and disordered branching events. \textbf{(c)} The instantaneous speeds averaged along the crack front corresponding to cases \textbf{(a)}  (thin red line) and \textbf{(b)} (thin blue line). Thick lines are the corresponding average velocities $v$ along a moving time window of size 20. The velocity of quasi bidimensional pattern \textbf{(a)} is significantly larger than that of 3D pattern \textbf{(b)}. The decrease observed at the end of the signal corresponds to a transition to 3D localized branching events and further evolution of the crack propagation leads to a pattern similar to \textbf{(b)}. \textbf{(d)} Average velocity of the crack front $v$ as function of $G/G_0$, the ratio of the available elastic  energy and the fracture energy. ($\Box$) correspond to results of 2D simulations~\cite{Henry08} and ($\bullet$) result from actual 3D simulations.}
\label{sensitivity_in_cond}
\end{figure}

When the crack speed is further increased, one observes a large variety of
patterns that appear to be due to the sensitivity to initial conditions and to
disorder. Indeed, in some situations, branching events  spread through the
whole width of the sample and give rise to a branching pattern that is very
similar to the one observed in two dimensions (see
Fig.~\ref{sensitivity_in_cond}(a)). Nonetheless, this is mostly observed when
the thickness $T$ is small. In the case of thick samples ($T\gg w_\varphi$),
these quasi bi-dimensional structures are rarely observed and the most frequent
pattern is  an apparently disordered array of branching events with varying
amplitude and thickness (see Fig.~\ref{sensitivity_in_cond}(b)). Quasi
bi-dimensional patterns tend also to undergo a transition toward the fully
tri-dimensional ones after a sufficiently long crack propagation. When
considering the average velocity of cracks this difference in the nature of the
patterns (3D or 2D) translates into the fact that cracks with quasi
bidimensional structures  have  higher average velocities than cracks with
three dimensional branching patterns (see Fig.~\ref{sensitivity_in_cond}(c)).
This is in qualitative agreement with the fact that the total crack surface is
significantly higher in the case of three dimensional patterns~:  the
difference in crack surface between the samples of
Fig.~\ref{sensitivity_in_cond}(a) and Fig.~\ref{sensitivity_in_cond}(b) is
approximately 20\%. Indeed, this increase in crack surface can be seen as an
additional energy dissipation at the crack tip, leading to crack slow down.
Fig.~\ref{sensitivity_in_cond}(d) confirms that for a finite range of  $G/G_0$
corresponding to a regime where patterns similar to
Fig.~\ref{sensitivity_in_cond}(b) are formed, the velocity measured during 3D
simulations is significantly smaller than the one measured for 2D simulations.
The measurements of both the energy flux into the front of the moving crack at a given speed and the total surface
area created via the microbranching instability show that the instability is the main mechanism for
energy dissipation by a moving crack in brittle materials.

\begin{figure}
\includegraphics[width=0.4\textwidth]{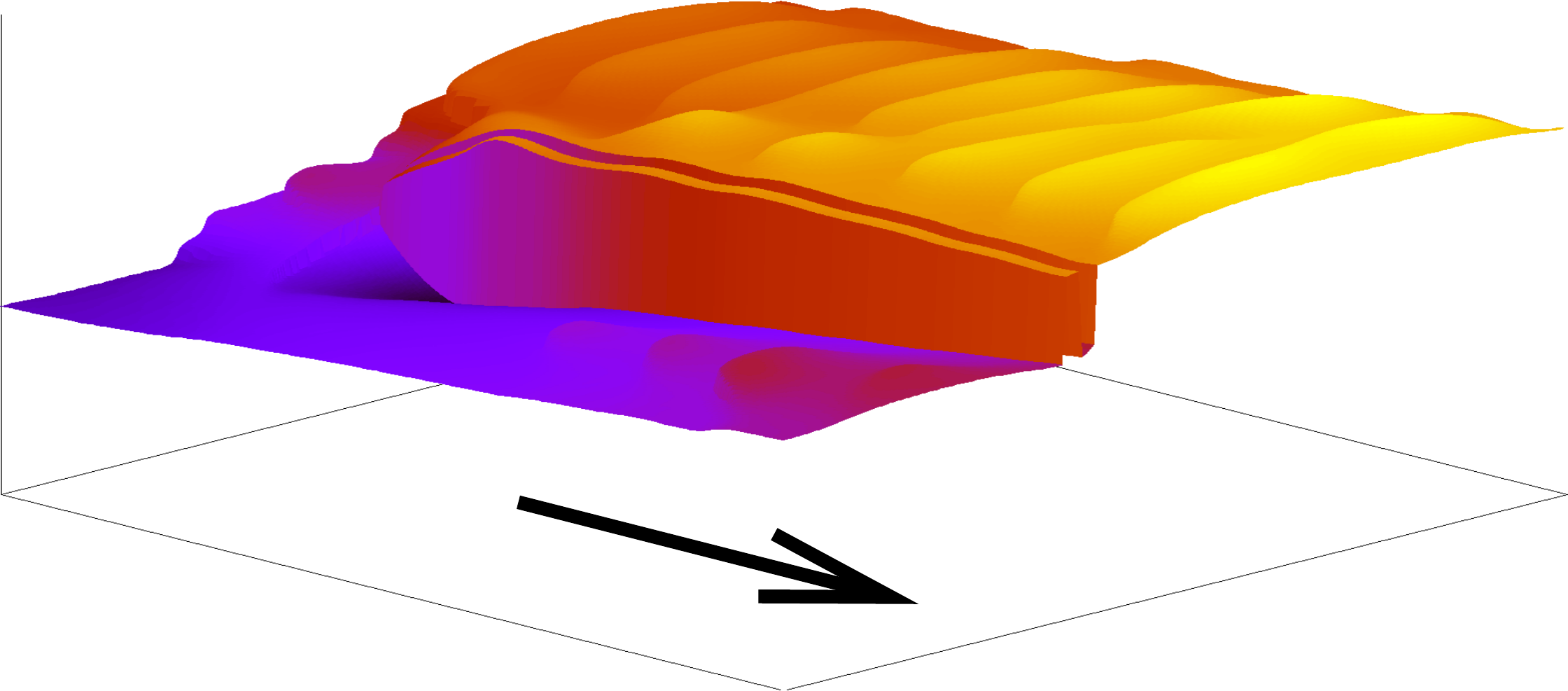}
\caption{\label{fig_echelons} A fracture surface with an echelon crack. The whole thickness of the material is shown and the secondary branches have been removed for clarity. Traction-free boundary conditions at $z=\pm T/2$ were used in the simulation. The arrow indicates the main direction of  crack propagation.}
\end{figure}

Finally, it should be mentioned that in some simulations the interplay of
disordered branching events lead to the birth of structures that are
reminiscent of the so called \textit{\'echelons cracks}. These are two cracks
propagating in the same direction but on different planes separated by a
discontinuity in the crack surface. Indeed, in some circumstances,  two
distinct planar cracks were propagating parallel to each other (see Fig.~\ref{fig_echelons}). The distance over which the echelons cracks persist is much larger for the case of traction-free surface boundary conditions than for the case of periodic boundary conditions. This difference can be attributed to the self interaction of a crack with itself that is induced by periodic boundary conditions. Hence under pure mode~I loading,  the localized branching event can induce  front fragmentation similarly to what is observed under mixed mode~(I+III) loading~\cite{Pons10,pollardseagall,baumberger2008}. 

The existence  of various morphologies within the same loading conditions leads
to the question of pattern selection by the system. As shown in
Fig.~\ref{sensitivity_in_cond} (a) and Fig.~\ref{fig_echelons}, one can see
that during the crack propagation the nature of the pattern can change. In both
cases, the \textit{transition} is first local and then spreads through the
sample as the crack propagates. Second, the introduction of quenched disorder
in the system, such as a random variation of the kinematic coefficient of the
phase field $\tau$,  favours strongly the three dimensional pattern.
	 
To conclude, the fast propagation of cracks was simulated in three dimensions.
As expected, the branching instability was observed and, even in a homogeneous
system, one could observe the birth of three dimensional patterns. These
patterns were induced by a sole instability mechanism~: a tip-splitting
instability of the crack front. Neither crack nucleation nor side-branching
(where a secondary crack nucleates at the surface of an existing crack) were
observed. The patterns found reproduce qualitatively various experimental
fractographic observations,{ despite the simplicity of the description of
the breaking process in the phase field model. This indicates that the
details of the breaking mechanism in the process zone  may play a
limited role in the \textit{formation} of fractographic patterns}. 
It is also fairly remarkable that for a given
parameter set, depending on the initial conditions and the position along the
crack propagation axis, different  patterns are observed. These results show
that the phase field model is a valuable tool to study  in depth  the
statistical aspects of dynamical and morphological instabilities of cracks
 and is a strong argument in favor of the use of diffuse interface approaches
to model crack propagation. For instance, the use of advanced computational techniques will allow to consider larger systems where finite size effects introduced by the phase field approach can be quantified.

\end{document}